\documentclass{aa}

\newcommand{\mdot}{\dot{M}}
\newcommand{\msun}{\mathrm{M}_{\odot}}
\newcommand{\spin}{a}
\newcommand{\mbh}{M_{\rm BH}}

\newcommand{\rhigh}{R_{\rm high}}

\graphicspath{{./}{figures/}}
\usepackage{xcolor}
\usepackage{setspace}
\usepackage{gensymb}

\begin{document}

\title{Deep Horizon: A machine learning network that recovers accreting black hole parameters}
\titlerunning{Deep Horizon}
\authorrunning{van der Gucht et al.}

\author{Jeffrey van der Gucht \inst{1}, Jordy Davelaar \inst{1,2}\fnmsep\thanks{j.davelaar@astro.ru.nl}, Luc Hendriks \inst{1}, Oliver Porth \inst{4,3}, Hector Olivares \inst{3}, Yosuke Mizuno \inst{3}, Christian M. Fromm \inst{3,5}, and Heino Falcke \inst{1}}

\institute{
           Department of Astrophysics/IMAPP, Radboud University, PO. Box 9010, 6500 GL  Nijmegen, The Netherlands
           \and
           Center for Computational Astrophysics, Flatiron Institute, 162 Fifth Avenue,  New York, NY 10010, USA
           \and
           Institut f\"ur Theoretische Physik, Max-von-Laue-Stra{\ss}e 1, D-60438 Frankfurt am Main, Germany
          \and
           Anton Pannekoek Instituut, Universiteit van Amsterdam P.O. Box 94249, 1090 GE Amsterdam, The Netherlands
           \and
           Max-Planck Institute for Radio Astronomy, Auf dem Huegel 69, D-53115 Bonn, Germany
          }

\date{Received 29/10 ; accepted 30/3}

\abstract
{The Event Horizon Telescope recently observed the first shadow of a black hole. Images like this can potentially be used to test or constrain theories of gravity and deepen the understanding in plasma physics at event horizon scales, which requires accurate parameter estimations.}
{In this work, we present {\tt Deep Horizon}, two convolutional deep neural networks that recover the physical parameters from images of black hole shadows. We investigate the effects of a limited telescope resolution and observations at higher frequencies.}
{We trained two convolutional deep neural networks on a large image library of simulated mock data. The first network is a Bayesian deep neural regression network and is used to recover the viewing angle $i$, and position angle, mass accretion rate $\mdot$, electron heating prescription $\rhigh$ and the black hole mass $\mbh$. The second network is a classification network that recovers the black hole spin $\spin$.}
{We find that with the current resolution of the Event Horizon Telescope, it is only possible to accurately recover a limited number of parameters of a static image, namely the mass and mass accretion rate. Since potential future space-based observing missions will operate at frequencies above 230 GHz, we also investigated the applicability of our network at a frequency of 690 GHz. The expected resolution of space-based missions is higher than the current resolution of the Event Horizon Telescope, and we show that {\tt Deep Horizon} can accurately recover the parameters of simulated observations with a comparable resolution to such missions.}
{} 

\keywords{black-hole physics, accretion, accretion disks, radiative transfer, methods: data analysis}

   \maketitle

\section{Introduction} \label{sec:intro}
In April 2019, the Event Horizon Telescope (EHT) collaboration released the first image of the shadow of a black hole \citep{EHT1, EHT2, EHT3, EHT4, EHT5, EHT6}. This image is direct evidence of the existence of black holes, a fundamental prediction of the general theory of relativity (GR) \citep{Schwarzschild1916, Kerr1963}.

In GR, astrophysical black holes are characterized by their mass, $\mbh$, and their spin, $\spin$ = $Jc/G\mbh^2$. In this equation, $J$ is the angular momentum of the black hole, $G$ is the gravitational constant, and $c$ is the speed of light. The size of the black hole is set by its event horizon, $R_{\rm h} = (1 + \sqrt{1 - \spin^2}) R_{\rm g}$, where $R_{\rm g} \equiv G\mbh/c^2$ is the gravitational radius. The event horizon defines a surface from within nothing can escape. The event horizon is gravitationally lensed, resulting in an effective angular size of $\theta = 2\sqrt{27} ~ R_{\rm g}/D$ for $a = 0$, where $D$ is the distance to the black hole. This lensed image is known as the shadow \citep{Falcke:1999pj}. Although the scale of the observed shadow is on the order $\theta$, the exact size depends on both the emission model and GR effects, such as spacetime rotation. Therefore, models of the accretion flow around black holes are needed to interpret the results of the EHT \citep{EHT5}.

The EHT array consists of eight telescopes positioned all around the globe, resulting in an effective resolution of $\sim 20$ microarcseconds ($\mu$as) when operating at 1.3 mm \citep{EHT2}. With this effective resolution, the EHT resolved the shadow of the black hole M87$^*$,  that is, a supermassive black hole (SMBH) in the nucleus of Messier 87. The distance to this SMBH is 16.8 $\pm$ 0.8 Mpc \citep{2010A&A...524A..71B, 2018ApJ...856..126C} and the mass is $6.5 \pm 0.7 \times 10^{9} M_\odot$ \citep{EHT6}. This mass estimate is computed from the observed angular size on the sky of 42 $\pm$ 3 $\mu$as \citep{EHT1}.

The size of the Earth limits the resolution of the EHT. Furthermore, the EHT only sparsely samples the Fourier domain of the image ($uv$-plane) \citep{EHT2}, owing to the limited amount of suitable millimeter Very Long Baseline Interferometry (VLBI) telescope sites. Increasing the amount of coverage in the $uv$-plane increases the quality of the image. Both of these limitations are mitigated by switching to space-based VLBI (SVLBI). Furthermore, SVLBI would remove atmospheric corruption and allow for longer baselines and higher frequencies. Therfore, SVLBI allows for higher resolutions and improved image quality, compared to ground-based VLBI. There are several studies of future SVLBI missions that observe the shadow of a black hole \citep{Palumbo2018, Fish2019, Roelofs2019}. \cite{Roelofs2019} report simulations of future SVLBI measurements of the black hole shadow of Sagittarius A$^*$, the SMBH in our galaxy, up to a frequency of 690 GHz. Their setup has baselines up to 60 G$\lambda$, resulting in a resolution of 4 $\mu$as after several months of observations. Density fluctuations in the interstellar medium electrons cause phase fluctuations in the incoming plane wave, resulting in scattering of the radio wave \citep{Narayan1989, Goodman1989, Johnson2015}. At a frequency of 690 GHz, there is less interstellar scattering \citep{Bower2006, Roelofs2019}, and the measured emission originates from closer to the event horizon as compared to the EHT observations.

The image of a black hole shadow can be used to test and constrain theories of gravity \citep{Johannsen2010, Psaltis2015, Goddi2017, EHT6, Mizuno2018}, but this requires accurate parameter estimations. Previous studies of M87$^*$ often use general relativistic magnetohydrodynamical (GRMHD) simulations to model the accretion flow \citep{Dexter2012, Moscibrodzka2016, Moscibrodzka2017, Ryan2018, Davelaar2018, Chael2019, Davelaar2019}. These studies fit their models to the observed spectra, resulting in constraints on the model parameters. In \cite{EHT5, EHT6}, the model parameters are constrained by fitting GRMHD models to the image of M87$^*$. The appearance of the black hole shadow in the image is determined by the parameters and can, therefore, be recovered directly from the image. In the case of EHT \citep{EHT5,EHT6}, GRMHD models were compared with the data. The models are either MAD or SANE, and include five spin values. The image library was then constructed based on these GRMHD models by performing general relativistic ray-tracing (GRRT) simulation for six values of the temperature ratio of electrons to protons inside the accretion disk, parametrized by $\rhigh$ \citep{Moscibrodzka2016, Moscibrodzka2017}, one mass, and two inclinations. The images were scaled in post-processing to test for other masses as well (e.g., not generated directly with ray-tracing codes). The current scoring of the GRMHD/GRRT images is conducted either via the single snapshot method (SSM; in \cite{EHT5})  or via the average image scoring (AIS; in \cite{EHT6}). Both approaches are performed in Fourier space, and a $\chi^2$ between the data and the model is computed using the visibility amplitude and closure phase.  During the fitting, the model images are rescaled (in flux density), rotated, and stretched (changing the mass to distance ratio). Currently, two pipelines--- {\tt THEMIS} using Markov chain Monte Carlo (MCMC)\citep{Broderick2020} and {\tt GENA,} using evolutionary algorithms\citep{Fromm2019}---are used to perform the fitting of the GRRT images. These pipelines require either MCMC steps, in the case of {\tt THEMIS,} or generations, in the case of {\tt GENA,} to provide matching between the data and the images. The results from this comparison, however, show that comparing images to the data results in almost all models fitting the EHT 2017 data (see Table 2 column 4 in \cite{EHT5}); models are mainly rejected based on uncertain measurements of the jet power of M87 at larger wavelengths (and scales). To both improve the extracting of black hole parameters and to decrease computational needs, we performed a proof of concept to use machine learning for this vital task. In recent years, machine learning algorithms have shown to be efficient and accurate in various fields of astrophysics, including in galaxy classification \citep{Odewahn1992, Weir1995, Suchkov2005, Ball2006, Vasconcellos2011, Fadely2012, Sevilla-Noarbe2015, Kim2015, Kim2017, Lukic2017}, gravitational wave parameter analysis \citep{George2018, Shen2019, Fan2019}, asteroseismology \citep{Bellinger2016, Hon2017, Hendriks2018}, and gravitational lensing effects \citep{Hezaveh2017, Levasseur2017, Petrillo2017, Jacobs2017}.

In this paper, we present {\tt Deep Horizon}, two Bayesian convolutional deep neural networks that can accurately recover the input parameters of an image of the shadow of an accreting black hole. This network was constructed as a proof of concept to investigate if deep neural networks are capable of obtaining black hole models parameters from horizon scale images. In this proof of concept, we focus on six parameters: the viewing angle with respect to the black hole spin axis, $i$, mass accretion rate, $\mdot$, temperature ratio of electrons to protons inside the accretion disk, mass of the black hole, $\mbh$, position angle (PA), and spin of the black hole, $\spin$. Our neural network also returns a Bayesian motivated uncertainty on the parameter estimations of the first five parameters mentioned above. We use synthetic images to train and test our neural network, and we restrict ourselves to a single SMBH, M87$^*$, for which we adopt a distance $D = 16.4$ Mpc; this is slightly smaller, (by 2\%) than the value used in \cite{EHT5}, but since it is a general scale factor for the total emission recorded it does not affect the results of this proof of concept. The data are generated at two frequencies, $230$ GHz (EHT) and $690$ GHz (SVLBI). Furthermore, we investigate the effects of convolving our images with a Gaussian beam as an approximation of a limited telescope resolution \citep{EHT5}.

We organize the paper as follows: In section \ref{sec:methods} we describe our synthetic data generation and the machine learning methods. In section \ref{sec:results} we show the performance of {\tt Deep Horizon} on mock observations. In section \ref{sec:Discussion} we discuss our results and future improvements. In section \ref{sec:conclusion} we summarize our results.

\section{Methods} \label{sec:methods}
Machine learning is a data-driven approach that requires sufficiently large data sets to train the algorithm. In the problem treated in this paper, observational data are limited. Hence, we have to rely on current simulations to generate mock observations of the environment near a black hole. 

We generated two data sets, each consisting of 100.000 images. Only the frequency varies between the two sets. The images are computed by post-processing five different GRMHD simulations. The GRMHD data are generated with the Black Hole Accretion Code ({\tt BHAC}) \citep{Porth2017, Porth2019, Olivares2019} \footnote{ Publicly availble at https://bhac.science}, and the post-processing is done via the GRRT code {\tt RAPTOR} \citep{Bronzwaer2018}\footnote{ Publicly availble at https://github.com/tbronzwaer/raptor}.

\subsection{GRMHD simulations}\label{subsec:GRMHD}
The two relevant physical parameters of a GRMHD simulation are the spin $\spin$ and the absolute magnetic flux $\Phi$ through the horizon often used in the dimensionless form $\phi = \Phi/\sqrt{\dot{M}R_{\rm g}^2 c}$ \citep{Tchekhovskoy2011, Porth2019}. In this paper, we only consider standard and normal evolution (SANE; \citealt{Narayan2012}) models with $\phi \sim 1,$ and we use models with a spin of $\spin$ = 0, $\pm$ 0.5 and $\pm$ 0.9375. These simulations are part of the simulation library that is used in \cite{EHT5} and are initialized with a weakly magnetized Fishbone-Moncrief torus \citep{Fishbone1976} in orbit around the black hole. The thermal pressure is perturbed with white noise to initialize the magnetorotational instability (MRI). The MRI causes angular momentum to be transported, triggering accretion onto the black hole. The differential rotation of the spacetime and magnetic field lines causes a magnetized jet to launch.

\subsection{ GRRT simulations} \label{subsec:RAPTOR}
To calculate mock observations, we post-processed the GRMHD data with the GRRT code {\tt RAPTOR}. This code calculates the flux density map at a given frequency by computing null geodesics, starting from a virtual camera, and simultaneously performing radiative transport calculations. We used emission and absorption coefficients for thermal synchrotron emission. The RAPTOR code used the ``fast light'' paradigm, where the simulation is frozen with respect to the elapsed photon time, which is equivalent to infinite light speed. We computed images at 230 GHz and 690 GHz. We used a camera with a field of view of ($0.1 \times 0.1$) milli-arcseconds$^2$ and generated images at ($128 \times 128$) pixels.

The GRMHD simulations are scale-free. Therefore, we had to convert the GRMHD variables from code units to centimeter-gram-second (cgs) units. This is done by defining the simulation length unit $\mathcal{L} = R_{\rm g}$, the simulation time unit $\mathcal{T} = R_{\rm g}/c$, and the simulation mass unit $\mathcal{M}$, where $\mathcal{M}$ sets the density in the accretion flow. The dimensionless accretion rate $\mdot_{\rm sim}$ can be converted into the accretion rate in cgs units by $\mdot = \mdot_{\rm sim} \mathcal{M}/\mathcal{T}$. The variables $\mbh$ and $\mdot$ are varied in our data generation. 

The GRMHD simulation does not evolve the radiatively important electrons. We used a parametrization for the plasma variables, which is based on the assumption that the proton-to-electron coupling depends on plasma magnetization \citep{Moscibrodzka2016, Moscibrodzka2017,Davelaar2018,EHT5,Davelaar2019}. This coupling is described by the following formula:

\begin{eqnarray}
    \frac{T_{\rm p}}{T_{\rm e}} = R_{\rm low}\frac{1}{1+\beta^2} + \rhigh \frac{\beta^2}{1+\beta^2} \,,
\end{eqnarray}

\noindent where $\beta = \frac{P_{\rm gas}}{P_{\rm mag}}$ is the ratio of the gas pressure, $P_{\rm gas}$, to the magnetic field pressure, $P_{\rm mag} = B^2/2$, where $B$ is the magnetic field strength. In the limit of $\beta \ll 1$, the temperature ratio asymptotically approaches $T_{\rm P}/T_{\rm e} \rightarrow R_{\rm low}$, while in the limit of $\beta \gg 1$ the temperature ratio asymptotically approaches $T_P/T_e \rightarrow \rhigh$. We set $R_{\rm low}$ to 1 and we vary $\rhigh$. We varied the viewing angle, which is defined as the angle between the observer and the black hole spin axis. Finally, we overlaid our images with a circular mask and rotated them to change the PA, the projected angle between the image plane, and the black hole spin axis.

The GRMHD simulations were run up to $t_{\rm final} = 10.000$ $\mathcal{T}$ consisting of 1.000 snapshots with an interval of 10 $\mathcal{T}$.  The correlation time of the image is $\sim$ 50 $\mathcal{T}$. In our data set, we used the last 100 snapshots of every spin value to capture the time evolution of the accretion flow. In these snapshots, the system is well evolved and the accretion flow has reached a quasi-steady state. We prevented our network from overfitting to single snapshots by randomly selecting ten snapshots as a validation set and training on the other snapshots. For each of these snapshots, we computed 200 images. Except for the spin, all parameters are randomly picked from a uniform distribution between the parameter ranges given in table \ref{parameter_ranges}. This ensures that there is no overlap between the training and validation sets. The mass prior is set such that it includes the one sigma range of the reported mass values of $3.5^{+0.9}_{-0.7}\times10^9 M_\odot$and $6.6\pm0.4\times10^9 M_\odot$ by \cite{gebhardt2011} and \cite{Walsh2013}.

Each parameter affects the image morphology differently. The viewing angle and PA influence the position of the jet and the asymmetry in the image. The density of the accretion flow determines the observed integrated flux and is related to the mass accretion rate $\mdot$. The size of the shadow is predominantly determined by the black hole mass $\mbh$. The value of $\rhigh$ is related to the region the emission originates from, where a low value of $\rhigh$ corresponds to a high concentration of emission in the disk and a high value corresponds to emission that predominantly originates from the jet. The black hole spin influences the geometry around the black hole and therefore the shape of the shadow and the asymmetry of the image. We show these effects in section \ref{results_image_libs}. A more detailed discussion of the effects of these parameters on the images can be found in \cite{EHT5}.

The viewing angle is sampled between $15^\circ$ and $25^\circ$ \citep{Walker2018}, the PA between $0^\circ$ and $360^\circ$, $\rhigh$ between 1 and 100, and $\mbh$ between $2 \times 10^9$ $\msun$ and $8 \times 10^9$ $\msun$ \citep{2011ApJ...729..119G, 2013ApJ...770...86W}. The mass accretion rate depends on the black hole mass, $\mdot \propto \mathcal{M}/\mbh$. To determine the prior of the mass accretion rate $\dot{M}$ for a fixed $M_{\rm bh}=6.5\times10^9 M_\odot$, we manually fit to $1$ Jy \citep{2015ApJ...807..150A} at 230 GHz for $R_{\rm high}=1$ and $R_{\rm high}=100$ for every spin case at a mass of $M=6.5\times10^9 M_\odot$. The resulting range of priors is then extended with one order of magnitude to increase the scope of the training data and ensure the inclusion of the M87* flux measurement. The resulting range of $\mdot$ is between $2 \times 10^{-6}$ solar masses per year ($\msun/{\rm yr}$) and $0.01~ \msun/{\rm yr}$ \footnote{Or between $1.5 \times 10^{-8} \dot{M}_{\rm Edd}$ and $7.3 \times 10^{-5} \dot{M}_{\rm Edd}$, where $\dot{M}_{\rm Edd}$ is the Eddington accretion rate for a black hole with mass $M=6.5 \times 10^9 \msun$.}. These parameter ranges are summarized in table \ref{parameter_ranges}. We sampled all parameters linearly, except $\mdot$, which is sampled logarithmically because it covers a large range over multiple orders of magnitude. We predicted log($\mdot$) and converted this back to the original value after training in order to prevent from biasing our machine learning network. We applied a min-max normalization to all parameters. Finally, we convolved our images at 230 GHz with Gaussian beams of 5, 10, and 20 $\mu$as. The latter is the current nominal resolution of the EHT; future arrays might correspond to higher resolutions when either 345 GHz is added to the array or large baselines are realized. This is done as an approximation of a limited telescope resolution. In this work, we ignored other telescope or measurement effects such as  limited $uv$-coverage. The expected SVLBI resolution is sufficiently high that we do not convolve the images generated at 690 GHz; we assumed resolution as achieved by a set of telescopes in low Earth orbits, which are capable of obtaining resolutions of approximately $3 \mu as$ (for more information see \cite{Roelofs2019}). As a result, we have five image libraries: one at 690 GHz and four at 230 GHz.

\begin{table}
\centering
\begin{tabular}{ll}
\hline
\hline
\label{parameter_ranges}

Parameter   & Range  \\ \hline
i           & [$15^\circ$, $25^\circ$] \\
$\mdot$     & [$2 \times 10^{-6} \msun/{\rm yr}$, $0.01 \msun/{\rm yr}$]\\
$\rhigh$    & [1, 100]  \\
M           & [2 x $10^9$ M$_\odot$, 8 x $10^9$ M$_\odot$]  \\
PA          & [$0^\circ$, $360^\circ$] \\
\hline
\spin       & 0,$\pm$ 0.5, $\pm$ 0.9375 \\ \hline

\end{tabular}
\caption{Model parameters. The parameter ranges used during data generation. The first five parameters are simulated continuously within the given range. The spin parameter $\spin$ is only simulated at five values.}
\end{table}

\subsection{Neural networks}\label{subsec:CNN}
Recent technological developments led to advancements in the fields of deep learning and computer vision \citep{Krizhevsky2012, Zeiler2013, Simonyan2014,Szegedy2014, He2015}. Computer vision using neural networks is typically done by training convolutional neural networks (CNNs) \citep{lecun1998, Krizhevsky2012}. We trained two CNNs in this work: a   Bayesian regression neural network and a classification neural network. 

Bayesian neural networks (BNNs) return a parameter estimation and a Bayesian motivated uncertainty \citep{MacKay1992, Gal2016, Kendall2017}. A BNN predicts two types of uncertainties: the aleatoric and the epistemic uncertainty \citep{Kiureghian2009, Kendall2017}. The aleatoric uncertainty is associated with corruption in the data, for example due to a limited resolution, whereas the epistemic uncertainty is related to uncertainty in the model parameters, for example due to an insufficient amount of data or training. Although there are many types of uncertainties, they are generally categorized as either aleatoric or epistemic \citep{Kiureghian2009}. Recent works have developed a fast and efficient method of approximating these uncertainties in machine learning \citep{Gal2015, Gal2015b, Kendall2017}. A neural network is trained by optimizing a loss function. By choosing a Gaussian log-likelihood loss function, the network is also able to predict the aleatoric uncertainty. We split the final network layer, so it returns both a prediction and an aleatoric uncertainty. The epistemic uncertainty is obtained by using variational inference with a method called Monte Carlo dropout (MCD). In MCD, dropout layers \citep{JMLR:v15:srivastava14a} are included after every weighted layer in the network. These layers have a fixed probability, the dropout rate, to turn off individual neurons for a single forward pass of the data through the network. The dropout rate is tuned such that the fraction of validation examples that lay within a certain confidence interval correspond to those of a normal distribution \citep{Levasseur2017}. Dropout layers add a random component to the network, which results in repeated predictions of the same image giving varying outcomes. The collection of repeated predictions is used to sample the posterior probability distribution, and the variance of this distribution gives the epistemic uncertainty. The uncertainty can be combined by adding the epistemic variance to the mean aleatoric variance. By sampling the posterior with N predictions, the combined uncertainty $\sigma$ is obtained by the following formula:

\begin{eqnarray}
    \sigma^2 = \frac{1}{N} \sum_{\rm i=1}^{\rm N} \hat{y}_{\rm i}^2 -(\frac{1}{N}\sum_{\rm i=1}^{\rm N}\hat{y}_{\rm i})^2 +\frac{1}{N}\sum_{\rm i=1}^{\rm N}\sigma_{\rm i, al}^2  \,,
\end{eqnarray}

\noindent where $\hat{y}$ is the network prediction and $\sigma_{\rm al}$ is the aleatoric uncertainty. For more details on this method, we refer to \cite{Kendall2017} or \cite{Levasseur2017}.

We used the regression BNN, hereafter network I, to predict the viewing angle $i$,  mass accretion rate $\mdot$, plasma parameter $\rhigh$, black hole mass $\mbh$ and t PA. We used the classification network, hereafter network II, to predict the black hole spin $\spin$. We chose a classification network for this parameter because we only have five distinct values in the training sets. Network I is trained with a negative Gaussian log-likelihood loss function, described by

\begin{equation}\label{neg_loss}
    -L = \sum_i \frac{1}{2 \hat{\sigma}_{\rm{i, al}}^2} \left| y_{\rm n,i} - \hat{y}_{\rm n,i} \right|^2 + \frac{1}{2} \log \hat{\sigma}_{\rm{i, al}}^2 \,, \\
\end{equation}

\noindent where $y_{\rm n,k}$ is the true value of the k'th parameter in the n'th image. Network II predicts the black hole spin $\spin$. We train this parameter with a categorical cross-entropy loss function \citep{hastie01statisticallearning}, described by

\begin{equation}\label{cat_cross_entr}
    L = - \sum_i^C y_{\rm o, i} \log p_{\rm o, i} \,, \\
\end{equation}

\noindent where $C$ is the number of classes, in our case five different spin values, $y_{\rm o, i}$ is a binary indicator whether the class label $i$ is the correct classification for observation $o$ and $p_{\rm o, i}$ is the predicted probability.

\begin{figure*}[!htb]
\centering
    \includegraphics[width=0.85\textwidth]{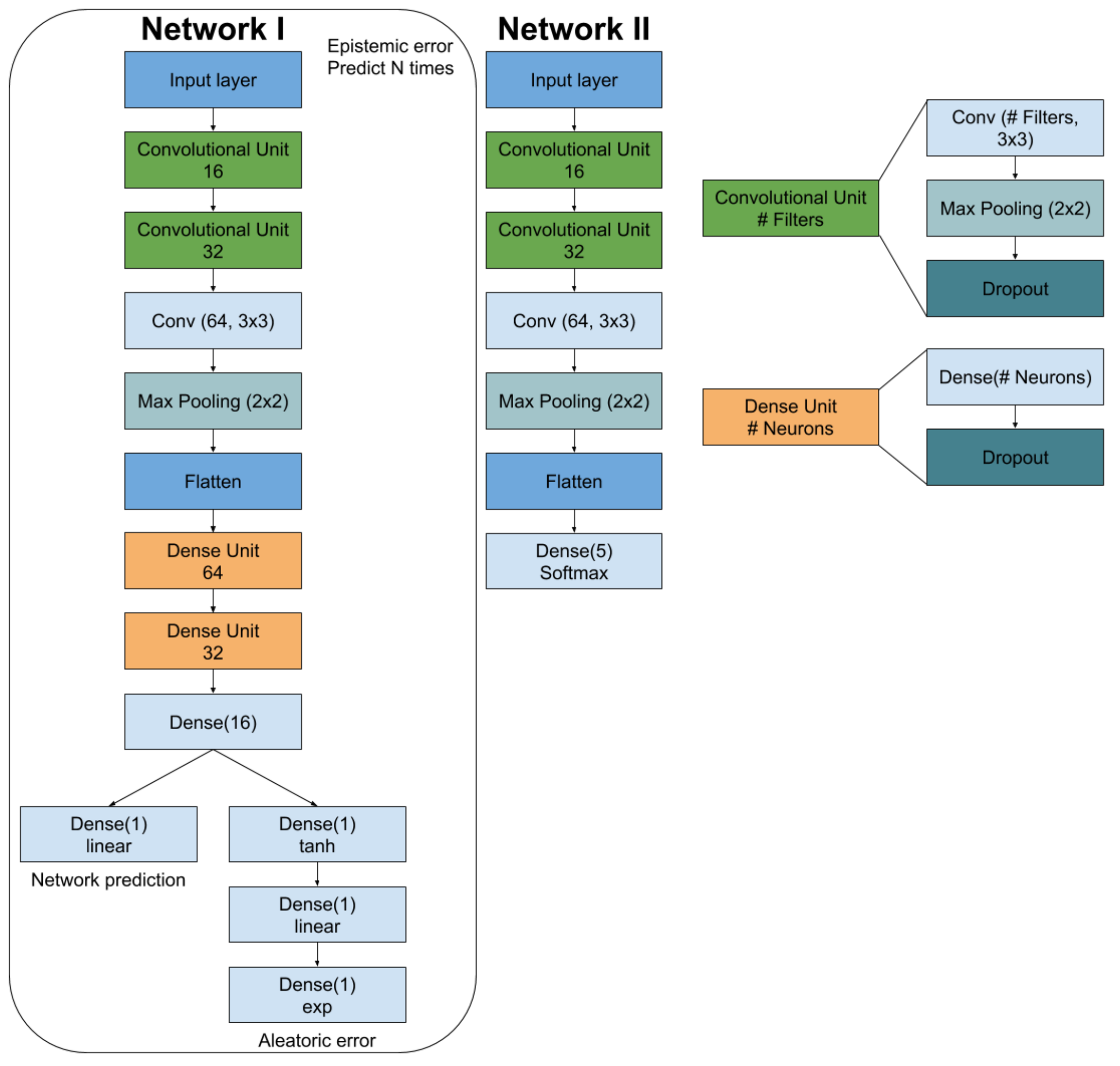}
       \caption{Network architecture. After the flatten layer, network I branch out into a dense network per parameter, resulting in five unique network arms, which allow for parameter-specific learning. The next time the arms branch out into a network prediction and an aleatoric uncertainty prediction. To capture the epistemic error of network I, we make N predictions on the same image to sample the network posterior. Both networks have ReLU activation functions unless stated otherwise.}\label{network}
\end{figure*}

The architecture of networks I and II can be found in figure \ref{network}. Network I and II have a similar architecture, except for the output layers, where we differentiate between the classification network and the regression network. The latter network has a further differentiation between the network prediction and the aleatoric uncertainty prediction. We split our image libraries into a training set consisting of 90.000 images and a validation set consisting of the remaining 10.000 images to prevent the network from overfitting on features in the training set. In all but the output layers, we used a rectified linear unit (ReLU) activation function, described by $f(x) = \max(0, x)$ \citep{conf/icml/NairH10}. We tuned the dropout rate as described in section \ref{subsec:CNN} and find 0.01 as our best value. The network is trained with the Adam optimizer \citep{2014arXiv1412.6980K} with Keras version 2.2.4 \citep{chollet2015keras} and TensorFlow version 1.11.0 \citep{tensorflow2015}. We set the initial learning rate on 0.001 and decreased it by a factor of 2 if the validation loss did not improve in two consecutive passes of the data through the network. We used a batch size of 32 during training. We used a random seed in TensorFlow of 1 to train our network, but we validated that our network can be trained independently of the chosen random seed. 

\section{Results} \label{sec:results}
\subsection{Image libraries}\label{results_image_libs}
\begin{figure*}[!htb]
\centering
    \includegraphics[width=0.85\textwidth]{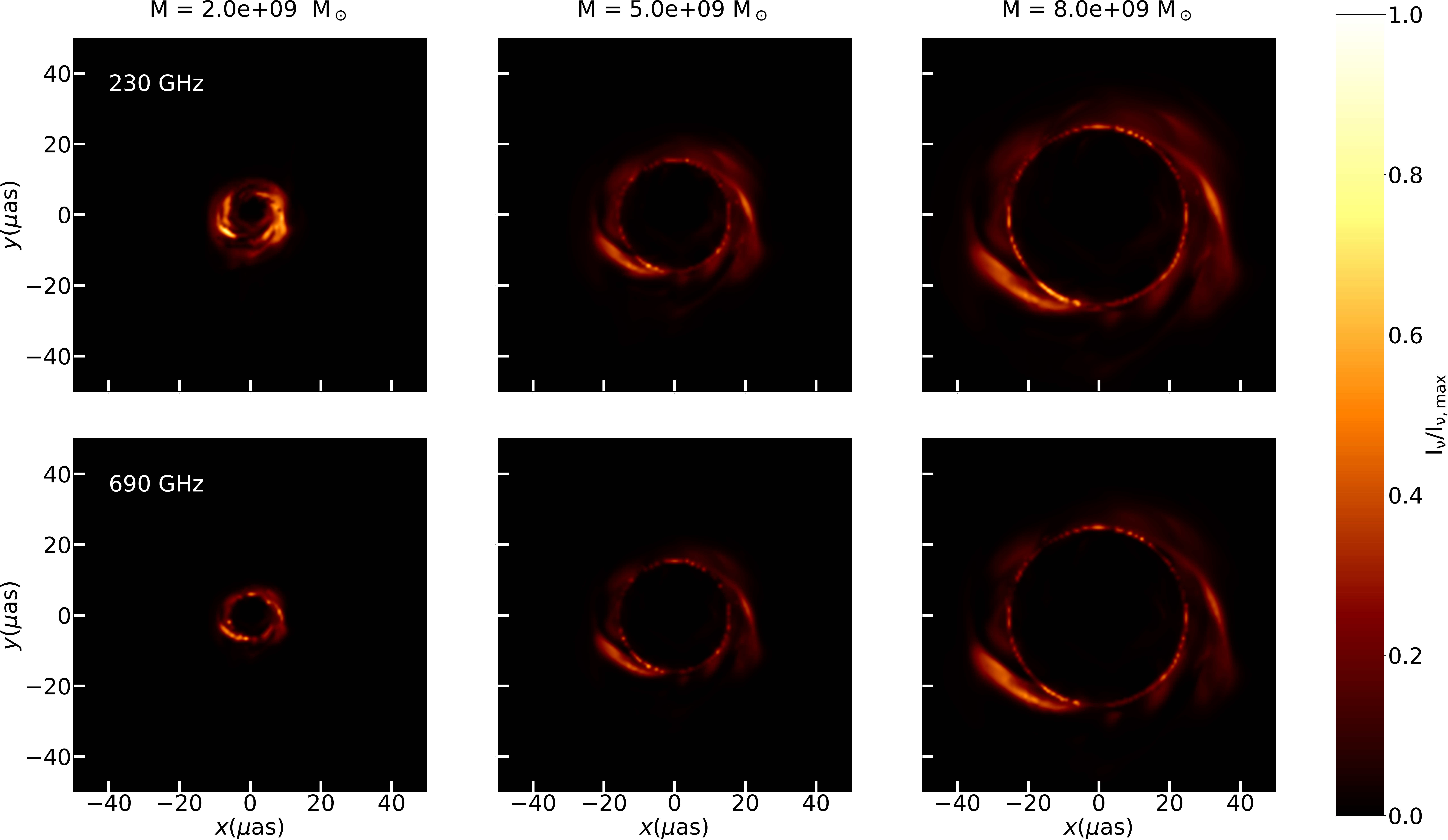}
       \caption{Single snapshot synthetic images. From left to right: 2.0 $\times 10^9$, 5.0 $\times 10^9$, and 8.0 $\times 10^9$ $\msun$. From top to bottom: 230 and 690 GHz. Images shown are representative images with a fixed flux of $F_{\rm 230 GHz}=0.5$ Jy for model parameters $a=-1/2$, $i=20^\circ$, $R_{\rm high}=50.0$, $\dot{M}=10^{-4} M_\odot/{\rm yr}$.}\label{lib_freq}
\end{figure*}

\begin{figure*}[!htb]
\centering
    \includegraphics[width=0.85\textwidth]{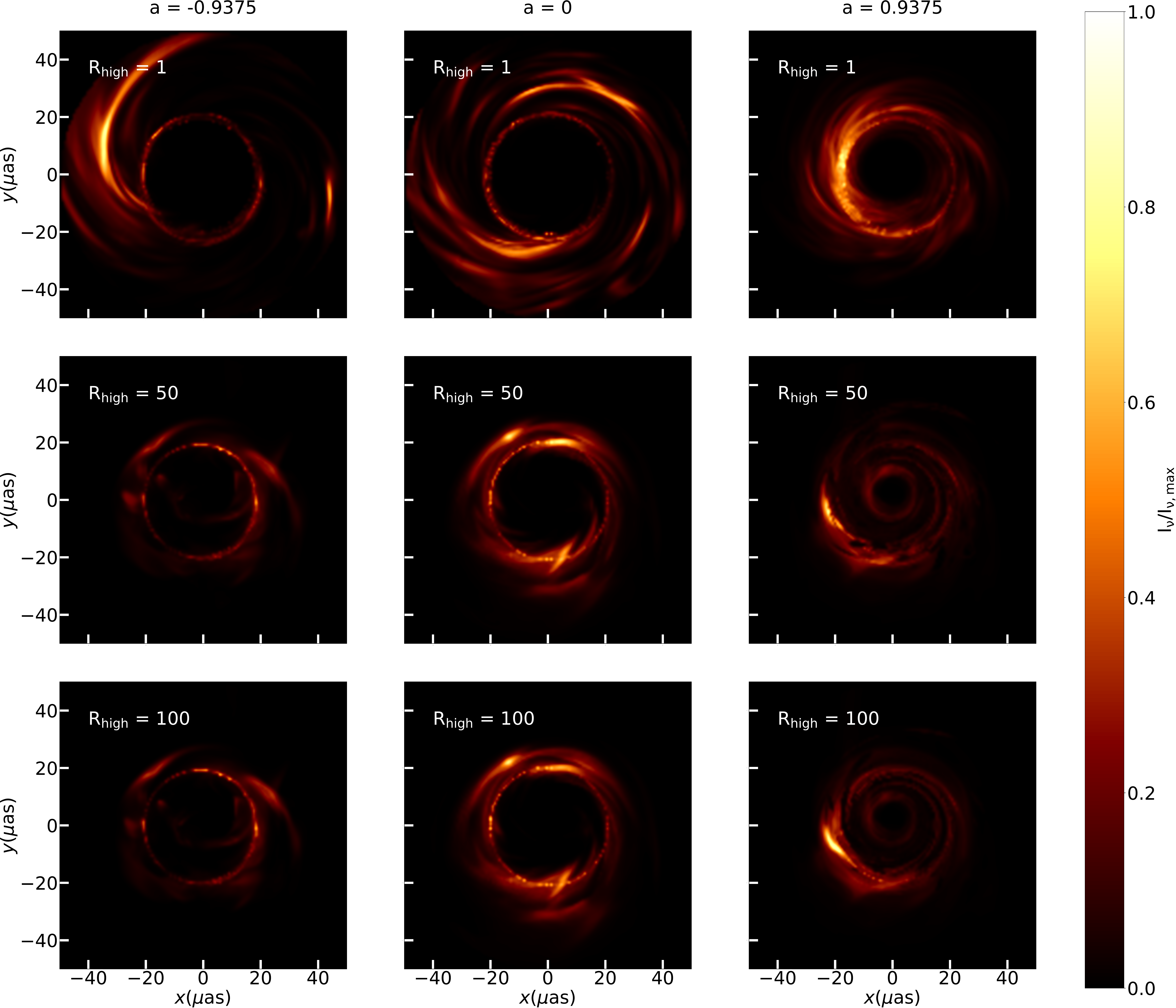}
       \caption{Single snapshot synthetic images at 230 GHz. From left to right: Spin values of -0.9375, 0 and 0.9375. From top to bottom: $\rhigh$ values of 1, 50 and 100. Images shown are representative images with a fixed flux of $F_{\rm 230 GHz}=0.5$ Jy for model parameters $i=20^\circ$ and $M=6.5\times10^{9} M_\odot$.}\label{lib_aR}
\end{figure*}

\begin{figure*}[!htb]
\centering
    \includegraphics[width=0.85\textwidth]{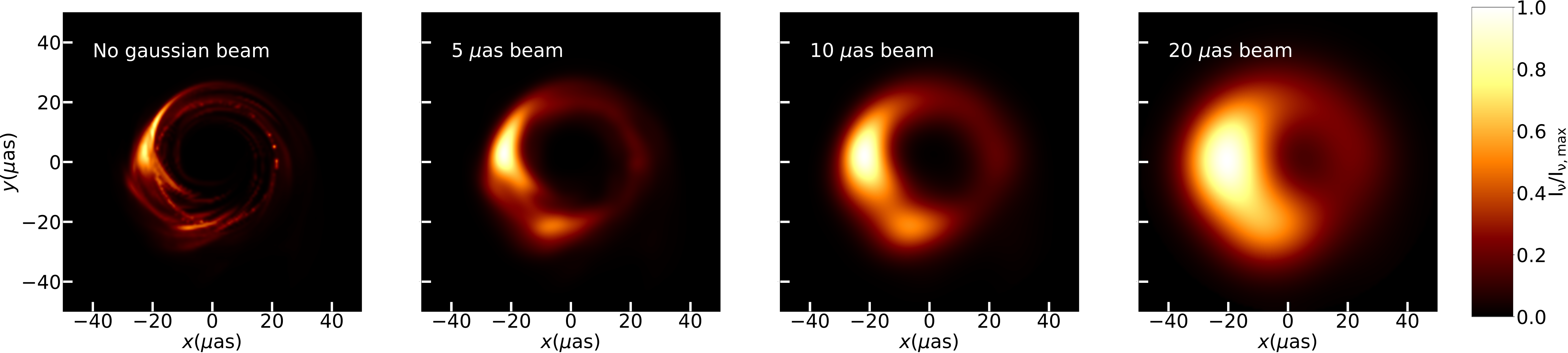}
       \caption{Single snapshot synthetic images. From left to right: No Gaussian beam and Gaussian beam widths of 5, 10, and 20 $\mu$as. Model used is identical to the bottom right model in Figure \ref{lib_aR}, with model parameters $a=0.9375$, $R_{\rm high}=100$, $i=20^\circ$ and $M=6.5\times10^{9} M_\odot$.}\label{lib_blur}
\end{figure*}

In figures \ref{lib_freq}, \ref{lib_aR}, and \ref{lib_blur}, we show example images of the image libraries. Every image is generated by running {\tt RAPTOR} with a unique set of parameters. The images show a central flux depression where the black hole is located, surrounded by a bright ring that coincides with the lensed emission ring \citep{Falcke2000,Gralla2019,Johnson2019,Narayan2019}. This ring scales with the black hole mass. Models with low values of $\rhigh$ show extended emission features that originate from the accretion disk of the black holes. Many of the small-scale features are lost when the Gaussian beam is applied. With our network, we investigate what minimal resolution is required to make reliable parameter estimations by varying the Gaussian beam widths. We demand that false predictions are reflected by smaller network confidence through larger uncertainties on the prediction. These results can be seen in subsection \ref{results_230GHz}. The effective size of the EHT array is limited by the size of the Earth. Therefore, large improvements in the resolution require higher frequencies. Planned extensions of the EHT to 345 GHz would improve the resolution by $\sim$ 40 $\%$ \citep{EHT2}. Further large improvements can be gained by switching to SVLBI. In subsection \ref{results_690GHz}, we show how our network performs at an SVLBI frequency of 690 GHz. The resolution of SVLBI experiments is expected to be sufficiently good enough to compare these experiments to the simulations without convolving them with a Gaussian beam.

\subsection{Event Horizon Telescope}\label{results_230GHz}
\begin{figure*}[!htb]
\centering
    \includegraphics[width=0.85\textwidth]{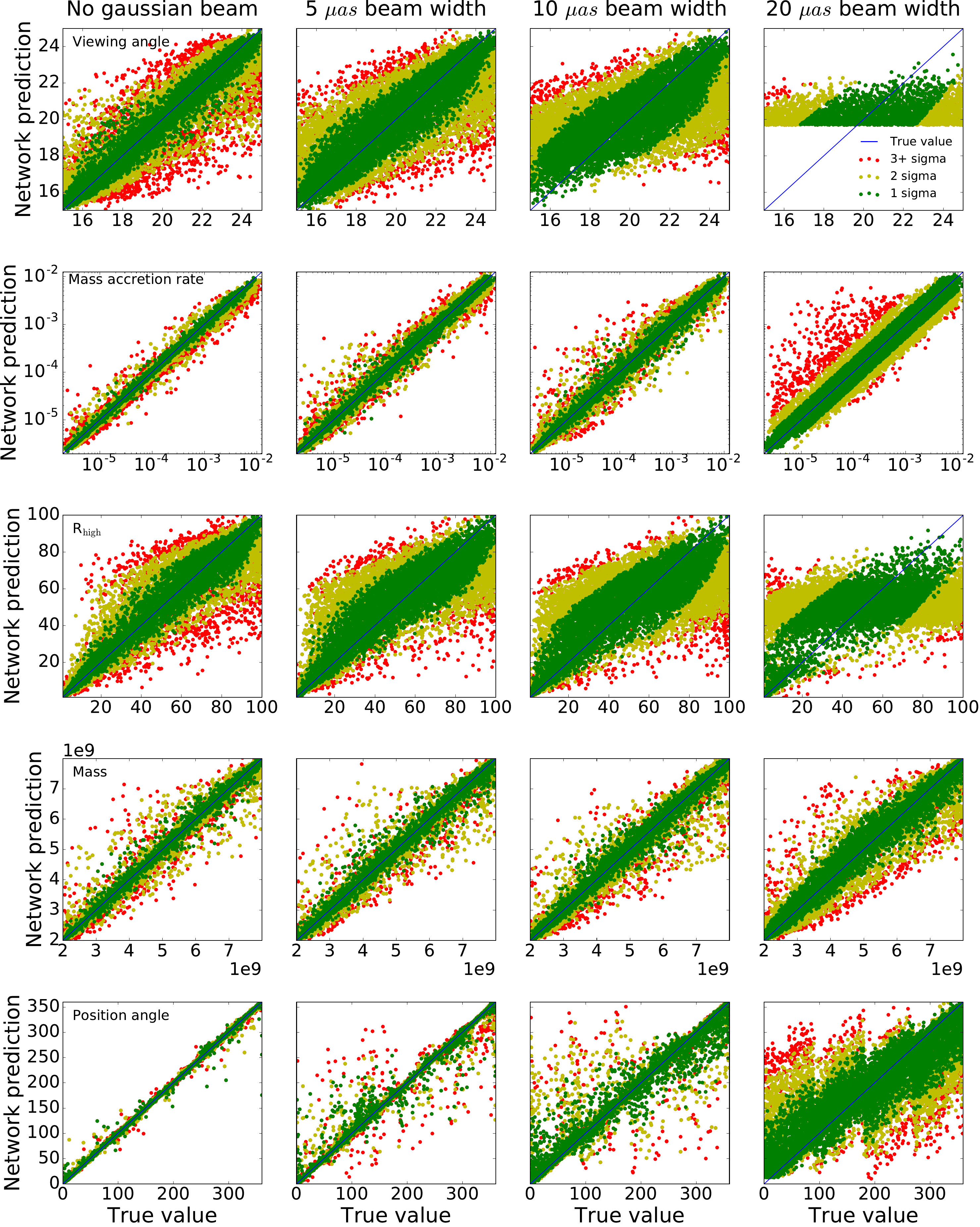}
       \caption{Predictions of Network I at 230 GHz. The predictions of the 10000 points in the validation set of network I at a frequency of 230 GHz with various amounts of Gaussian beam widths. The color coding is based on the deviation of the network prediction with respect to the true value, weighed by the network uncertainty. The red points indicate predictions that are correct within three or more $\sigma$. The units of the viewing angle and PA are degrees ($^\circ$) of the mass accretion rate solar masses per year ($\msun/{\rm yr}$) and the mass is expressed in solar masses ($\msun$). Left to right: No Gaussian beam and Gaussian beam widths of 5, 10, and 20 $\mu$as. Top to bottom: The viewing angle, mass accretion rate, $\rhigh$, mass, and PA. The sawtooth pattern in the PA at large Gaussian beams is discussed in section \ref{Subsec: Discussion1}}\label{blurred_regression}
\end{figure*}

\begin{figure*}[!htb]
\centering
    \includegraphics[width=0.85\textwidth]{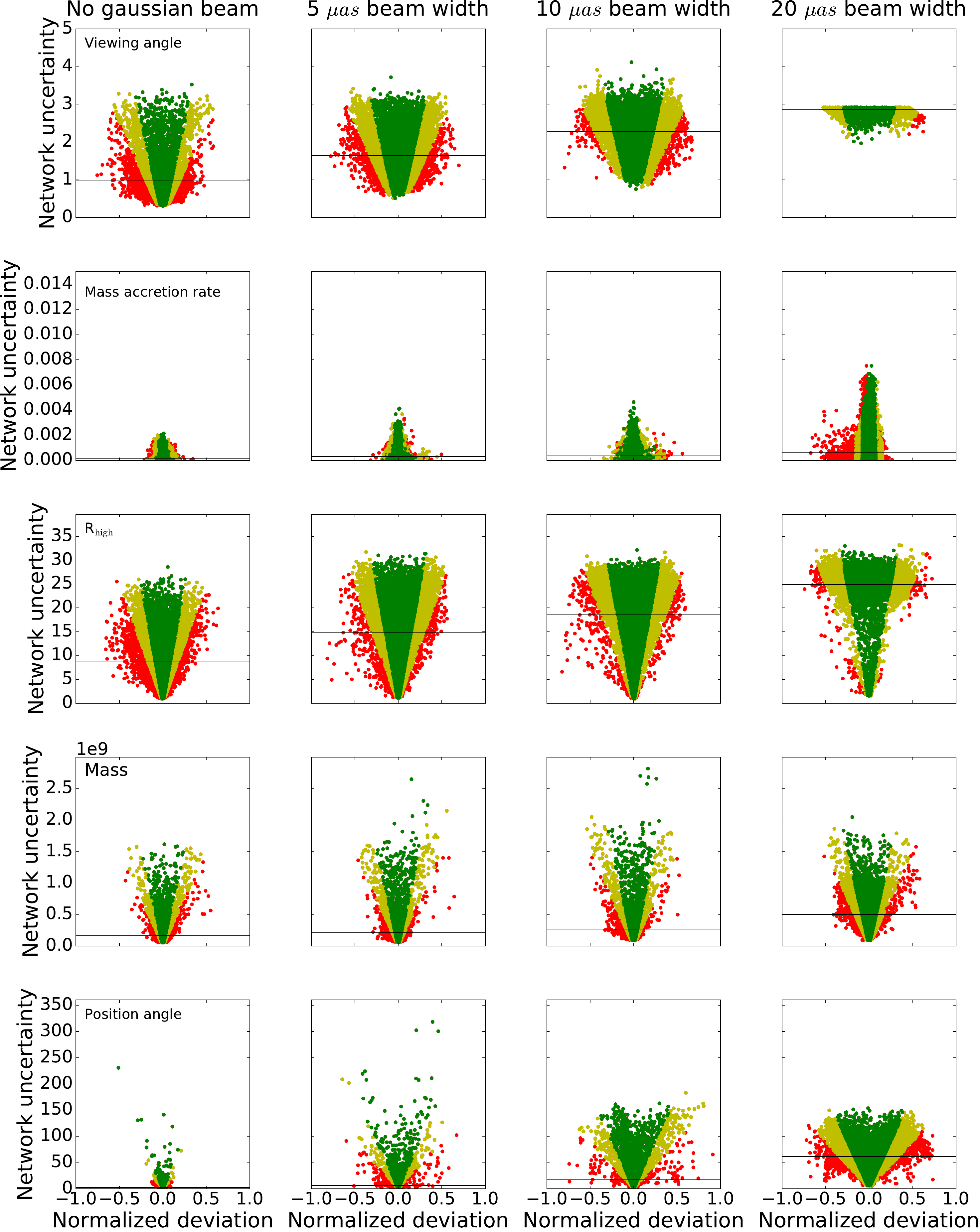}
       \caption{Network I uncertanties at 230 GHz. The network uncertainties as a function of the deviation of the network prediction with respect to the true value. The deviation is normalized to the parameter ranges, where a deviation of one corresponds to a maximally wrong network prediction. The color coding, units, and order of the figure are similar to figure \ref{blurred_regression}. The horizontal line indicates the mean predicted uncertainty.}
        \label{blurred_regression_errors}
\end{figure*}

We plot the network I predictions of the 230 GHz image libraries as a function of the true values in figure \ref{blurred_regression}. For each prediction, we determine the deviation for the correct prediction and weigh it with the uncertainty $\sigma$. This allows us to define how many points are predicted correctly within 1, 2, and 3 $\sigma$, which corresponds to approximately 68\%, 95\%, and 99\%. Without a Gaussian beam, the network can reliably predict the black hole parameters. A large scatter on the prediction indicates that the network is less confident, which is further reflected by a larger (mean) uncertainty. We show this uncertainty as a function of the deviation in figure \ref{blurred_regression_errors}. The horizontal line in this figure indicates the mean uncertainty. The values of the mean uncertainties can also be found in table \ref{mean_errors}. With an increasing beam width, the scatter in the uncertainty and the mean value of the uncertainty increase.

\begin{figure*}[!htb]
\centering
    \includegraphics[width=0.85\textwidth]{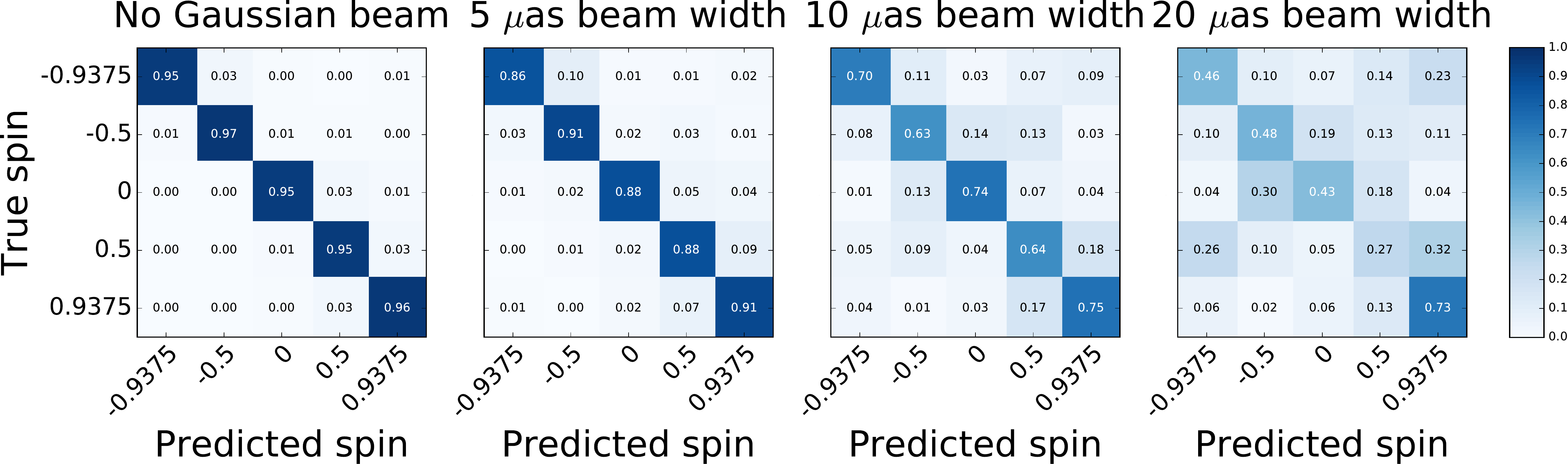}
       \caption{Network II predictions at 230 GHz. The predictions of network II at a frequency of 230 GHz. The y-axis of the confusion matrix represents the true labels and the x-axis represents the network predictions. The individual cells show the accuracy of a given combination. The five classes are equally represented, and the total validation set contains 10.000 images}
        \label{conf_matrix}
\end{figure*}

In figure \ref{conf_matrix}, we show a confusion matrix of the network II predictions. The neural network outputs a probability that an image belongs to a certain class and selects the class with the highest probability as network prediction. Without a Gaussian beam, the network has an accuracy of 95.9$\%$, where most of the false classifications are either one spin value higher or lower than the correct value. The five classes are equally represented, and the entire validation set we show in this work contains 10.000 images. Up to a Gaussian beam of 10 $\mu$as, the network can discriminate between the different spin values, resulting in high accuracy and minor deviations between the different classes. At 20 $\mu$as beam width, this is no longer true, and this results in many more than half of the predictions being bad and larger discrepancies between the classes. We further investigate the predictions by plotting the  receiver operating characteristic (ROC) curves and the corresponding area under the curves (AUCs). In a ROC plot, the true positive rate (\textsc{TPR}) is plotted versus the false positive rate (\textsc{FPR}) as a function of the classification threshold. These quantities are given in equation \ref{tpr_fpr}, where \textsc{TP} stands for the number of true positives (prediction and truth are both true), \textsc{P} stands for the number of true values in the data set, and \textsc{FN} stands for the number of false negatives (prediction is false, while truth is true), as follows:
\begin{equation}\label{tpr_fpr}
    \rm{TPR} = \rm{TP} / \rm{P} \\
    \rm{FPR} = \rm{FN} / \rm{P}
\end{equation}

The classification threshold is a minimal network probability that is required to belong to a certain class: with a threshold of zero, all spin values are compatible, whereas, with a threshold of one, only a perfect prediction is accepted. The AUC can be quantified with the integral defined as
\begin{equation}\label{auc}
    \rm{AUC} = \int_0^1 \rm{TPR}(\rm{FPR}^{-1}(x))dx.
\end{equation}

\begin{figure*}[!htb]
\centering
    \includegraphics[width=0.85\textwidth]{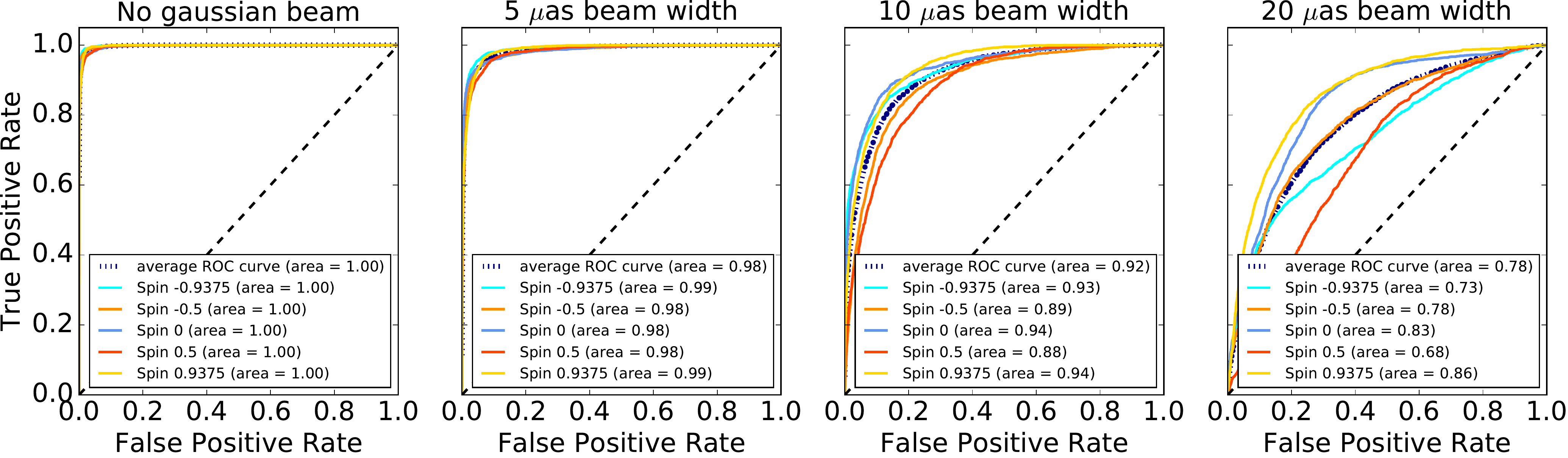}
       \caption{Receiver operating characteristic curve and AUCs at 230 GHz. The ROC curves and corresponding AUCs at a frequency of 230 GHz. Accurate classifiers have an AUC of 1, whereas inaccurate classifiers have an AUC of 0.5. There are no large discrepancies between different spin values.}
        \label{ROC}
\end{figure*}

In an accurate classifier, there is a high TPR at low FPR, which results in an AUC of 1. However, if the classifier cannot discriminate between the classes, they all are assigned approximately equal probabilities. Therefore, increasing the classification thresholds results in equally many true positives as false positives being accepted, which results in an AUC of 0.5. The ROC curves and the corresponding AUCs can be found in figure \ref{ROC}.

Up to a Gaussian beam width of 10 $\mu$as, {\tt Deep Horizon} reliably predicts the parameters of this study. With larger beams, the images are more alike and, therefore, harder to distinguish. This results in larger (mean) uncertainties and lower AUCs. Some parameters are heavily affected by this (e.g., the viewing angle and $\rhigh$), whereas the parameters that predominantly affect large-scale features such as $\mdot$ (total flux) and $\mbh$ (size of the shadow) are less affected by the Gaussian beam. For the viewing angle, $i$ the addition of a 20 $\mu$as blur results in almost all predictions of the network to be close to the mean of the training data (for $i$ at $20^{\degree}$). This is also reflected by the high uncertainty for all values, as can be seen in Fig. \ref{blurred_regression_errors}. The samples that result in network predictions, which are above the mean, show a majority of models with small black holes masses, high mass accretion rates, large $R_{\rm high}$ values, and larger viewing angles. These models have in common that they have a jet feature that potentially helps the network give a slight advantage to larger inclinations angles. 

\begin{table*}[!htb]
\centering
\label{mean_errors}
\begin{tabular}{lllll|l}
\hline
\hline
                                            & \multicolumn{4}{c|}{230 GHz}   & 690 GHz              \\ \hline
\multicolumn{1}{l|}{Beam width ($\mu$as)}   & 0                                 & 5  & 10  & 20 & 0            \\ \hline
\multicolumn{1}{l|}{$i$ ($^\circ$)}       & 1.0                  & 1.6                  & 2.3                   & 2.9                   & 1.0                  \\
\multicolumn{1}{l|}{$\mdot$ ($\msun/{\rm yr}$)} & 1.8 $\times 10^{-4}$ & 2.9 $\times 10^{-4}$ & 3.6 $\times 10^{-4}$  & 6.4 $\times 10^{-4}$  & 2.4 $\times 10^{-4}$ \\
\multicolumn{1}{l|}{$\rhigh$}             & 8.9                 & 14.8                 & 18.7                  & 24.9                  & 11.5                 \\
\multicolumn{1}{l|}{$\mbh (\msun)$}       & 1.6$\times 10^{8} $  & 2.1$\times 10^{8} $  & 2.7$\times 10^{8} $   & 5.0$\times 10^{8} $   & 1.5$\times 10^{8} $  \\
\multicolumn{1}{l|}{PA  ($^\circ$)}       & 2.9                  & 6.4                  & 17.2                  & 61.6                  & 4.5       \\ \hline

\end{tabular}
\caption{Mean uncertainties of network I. The individual predicted network uncertainties can be seen in figure \ref{blurred_regression_errors}. Left: 230 GHz. Right: Same as left but at a frequency of 690 GHz.}

\end{table*}

\subsection{Space VLBI}\label{results_690GHz}
\begin{figure*}
\centering
    \includegraphics[width=0.85\textwidth]{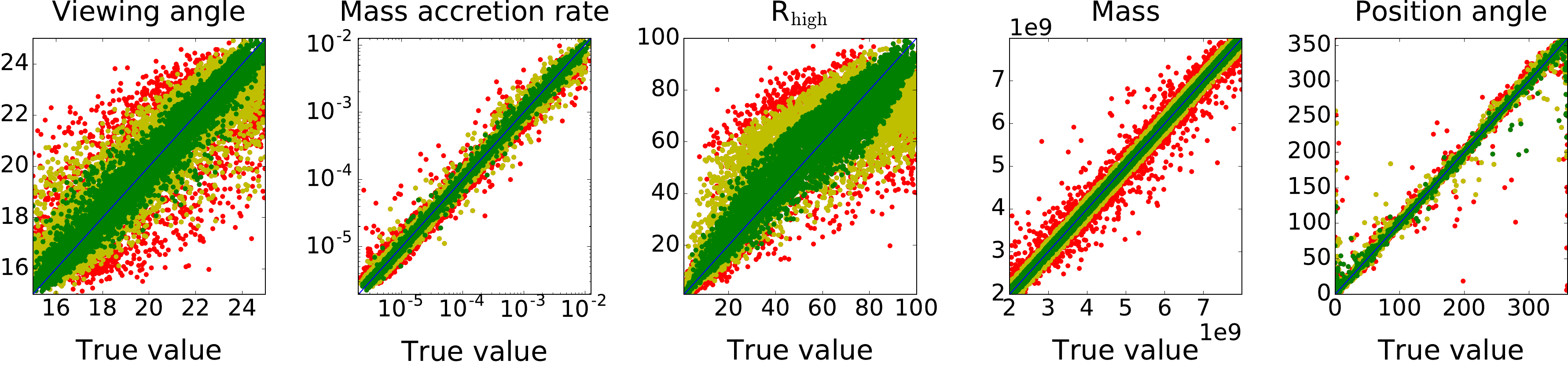}
       \caption{Network I prediction at 690 GHz. The predictions of network I at a frequency of 690 GHz, no Gaussian beam applied. The color coding and units are similar to figure \ref{blurred_regression}. Left to right: the viewing angle, mass accretion rate, $\rhigh$, mass, and PA.}
        \label{690_scatter}
\end{figure*}

\begin{figure}[!htb]
\centering
    \includegraphics[width=0.35\textwidth]{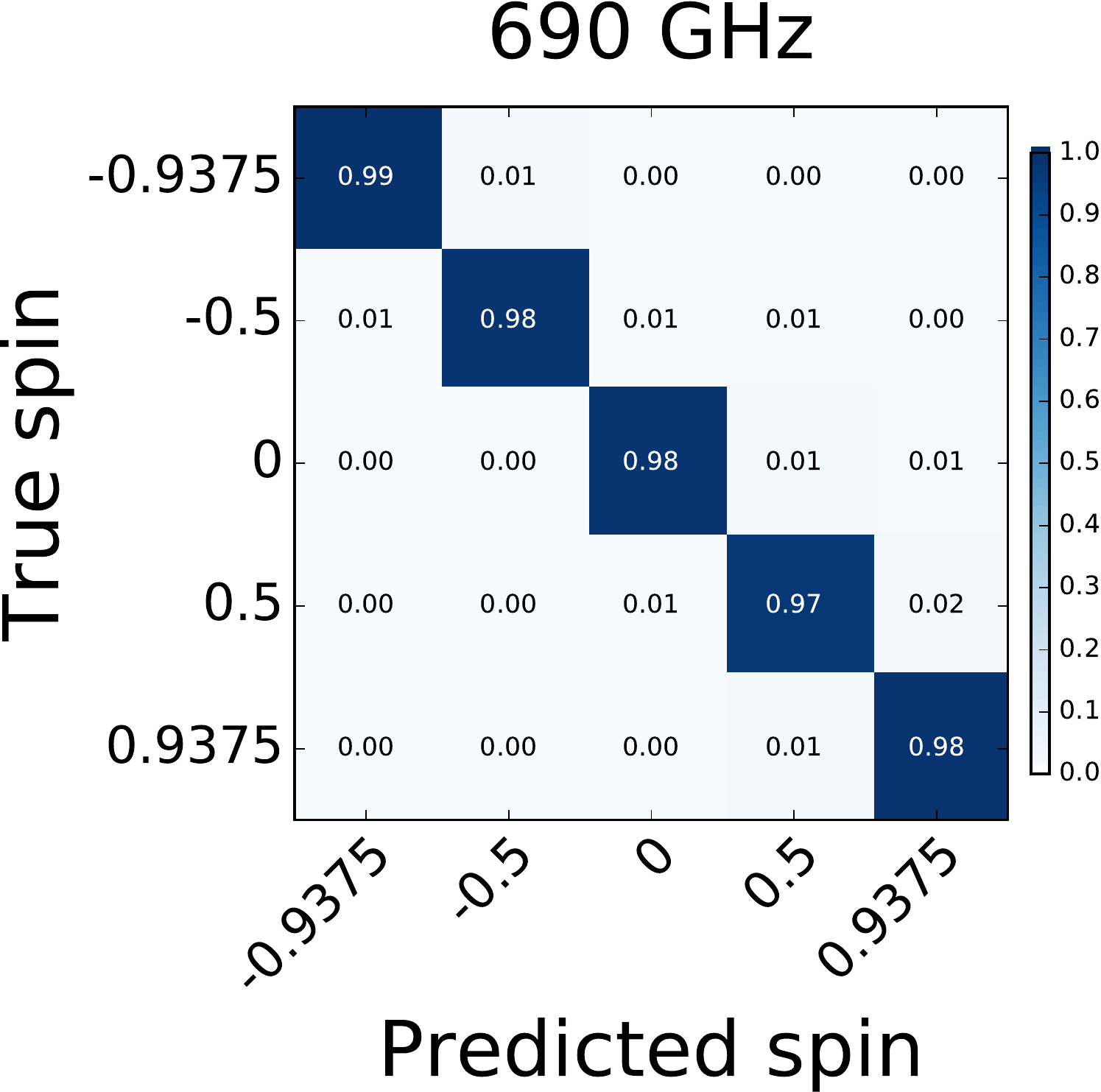}
        \includegraphics[width=0.35\textwidth]{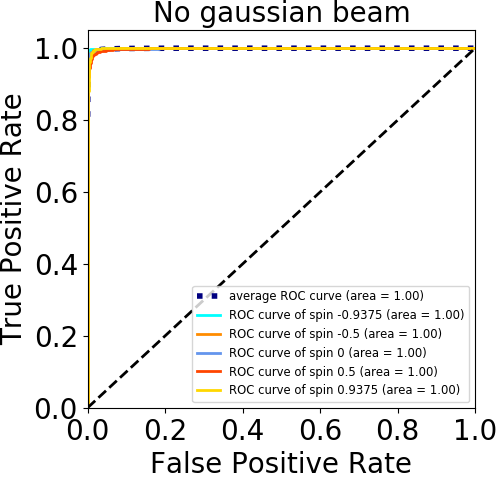}

       \caption{Network II predictions at 690 GHz. Top: Similar to figure \ref{conf_matrix} but at a frequency of 690 GHz.  Bottom: Similar to figure \ref{ROC} but now for 690 GHz. For both panels, no Gaussian beam is applied}
        \label{690_spin}
\end{figure}

We show the predictions of network I in figure \ref{690_scatter}, of network II in figure \ref{690_spin}, and the mean uncertainty per parameter in table \ref{mean_errors}. The performance at 690 GHz is comparable to 230 GHz without a Gaussian beam. The mass accretion rate, black hole mass, and PA have a relatively low amount of scattering in the network predictions, which is reflected by the relatively low mean uncertainties. The average accuracy over all spin values of network II is 98.1$\%$.

The expected data quality of SVLBI experiments allows for a direct comparison to mock data without convolving the images. We find that we can accurately recover all the parameters considered in this study. Therefore, future SVLBI missions would allow for more detailed measurements of SMBH systems.

\section{Discussion} \label{sec:Discussion}
\subsection{Network and data quality} \label{Subsec: Discussion1}
 In a machine learning algorithm, the training data are considered to be the ground truth. Therefore, the quality of the data set is important because any deviations of the simulation with respect to reality result in an unquantifiable uncertainty on the prediction. We have investigated the effects of increasing or decreasing the amount of training data and find that our algorithm is robust toward these changes. There are several ways we can further expand our image library to include more realistic mock observations. The first way is by including more parameters. In this proof of concept, we limit ourselves to SANE models for the accretion flow. However, the EHT observations are also in agreement with magnetically arrested disk (MAD; \citealt{Bisnovatyi1976, Narayan2003}) models for the accretion flow. We also only investigate emission models based on a thermal distribution function for the electrons. Possible alternatives that relax the commonly used thermal distribution functions are either $\kappa$-distribution function \citep{Davelaar2019} or power-law models \citep{Dexter2012}. Also, the choice of electron heating prescription limits the range of training data. Newly developed models include a secondary electron fluid that is evolved with the GRMHD simulation \citep{ressler2015,Ryan2018,chael2019b}, but these models highly depend on the choice for the underlying heating mechanism. In this proof of concept, we chose to use $R_{\rm low}=1$; this choice is identical to \cite{EHT5}. The value of $R_{\rm low}$ could, however, be smaller or lower than one, in general, the electron and proton fluid does not have to be in exact equilibrium in the jet sheath since electron heating in magnetized environments is shown to be more efficient \citep{howes2010,chandra2015,rowan2017}. To extend the predictive power of the network, a more diverse set of electron temperature models should be considered, for example, by including a larger range of $R_{\rm low}$ values. Another limitation of the training data can be found in the sampling of the spin parameter. Currently, only five spin simulations are present. The low sampling of the parameter space could result in overfitting on these examples when a classical regression network would be used. When applying our regression network to spin, large uncertainties are obtained owing to a limited amount of examples, which is expected for aleatoric uncertainties. To be able to perform regression on the spin, the number of values should be increased by at least a factor 2-4. Furthermore, we could also include images that are generated with theories beyond GR. Thereby, the algorithm could learn to recognize if an image is compatible with GR or one of the alternatives \citep{Mizuno2018}. Including these additional parameters is beyond the scope of this proof of concept, but we would like to investigate this in future studies. Another method to improve our data sets is by including realistic telescope effects. In our data sets, we approximate telescope resolution with a Gaussian beam but ignore effects such as thermal noise or telescope systematics. These effects are captured in {\tt SYMBA} \citep{Roelofs2019b} and the {\tt eht-imaging} \citep{Chael2018, Chael2019} Python package, which generate realistic synthetic data that can be reconstructed with imaging techniques. Improvements of the network could then go two ways, either by using reconstructed images based on synthetic data generated with GRMHD models or directly training the network on the visibility quantities in the synthetic data sets; these are two options that should be compared in follow-up works. Finally, we also ignore various constraints on the measurements such as spectral energy distribution fitting, dynamical measurements, and polarization. Including this information in future studies could further improve our network efficiency.

Figure \ref{blurred_regression} shows asymmetries in the network predictions of the PA with a large Gaussian beam. Upon inspection of the data, we see that in many of the inaccurately predicted points, the jet is not visible owing to the Gaussian beam. Instead, the Gaussian beam sometimes magnifies a local overabundance of plasma in the disk, which then shows up as if the jet is pointing in that direction. Furthermore, the loss function, as given in equation \ref{neg_loss}, does not capture the cyclicity of the parameter. Therefore, we introduce a bias in the PA that overestimates low PA values and underestimates high PA values. This causes the sawtooth pattern observed in figure \ref{blurred_regression}. In future works, we want to investigate the effects of modifying the loss function to remove this asymmetry. 

\subsection{Time evolution}
The environment near a black hole is a dynamic system. In our data generation, we use the last 100 snapshots of every spin value to capture these dynamics and prevent our network from overfitting on temporal features. We investigate the effects of the time evolution on our network by generating a new data set that has a sufficiently large temporal separation to be used as an independent test. We find no large discrepancies between the independent test set and our standard data sets, and therefore, conclude that our network is not overfitting the temporal correlations within the data.

\subsection{Comparison EHT}
In \cite{EHT6}, three independent algorithms are employed to quantify the size, orientation, and shape of the asymmetric ring structure found in the 2017 EHT observations. These three methods are geometric crescent model fitting, GRMHD model fitting, and image domain feature extraction. In this paper, we present a fourth independent method that can be used for parameter estimations. In this subsection, we discuss the error budget obtained in the EHT methods. We focus on the measurement of the angular size corresponding to a gravitational radius, $\theta_{\rm g} = R_{\rm g}/D$, which is used to find the black hole mass by folding in a distance measurement for M87.

The EHT reports three sources of uncertainty: a statistical uncertainty that corresponds to the width of the posterior, an observational uncertainty that corresponds to an incomplete ($u, v$) coverage, and unmodeled systematics and a theoretical uncertainty associated with the data being a single sample from a dynamic system. Of these three components, theoretical uncertainty is the largest component. Details on how the uncertainties are calculated per method are provided in the appendices of \cite{EHT6}. These individual contributions are further classified as systematic uncertainties due to the GRMHD calibration of the method and a statistical uncertainty originating from the angular diameter measurement. The average values over the different methods after folding in the distance measurement are $\sigma_{\rm sys}$ = 0.7 $\times 10^9 \msun$ and $\sigma_{\rm stat}$ = 0.2 $\times 10^9 \msun$.

Although the same GRMHD simulations are used by the EHT and in this work, these uncertainties do not describe the same uncertainties as  the uncertainties output by Deep Horizon because the methods to obtain these uncertainties are very different. Therefore, we cannot directly compare our network uncertainty to the uncertainty found within \cite{EHT6}. Furthermore, such a comparison would be beyond the scope of this paper because we do not test our network on the real image obtained by the EHT collaboration. However, we note that the mean uncertainties on the mass in table \ref{mean_errors} are of the same order of magnitude as the uncertainties found by the EHT. Although our method looks promising, further improvements and detailed method comparisons are required before we can apply this to observed data. One such comparison that requires further study is that, although our method differs from those described by the EHT, systematic correlations may remain as a result of the same underlying GRMHD simulations.

\section{Conclusions}\label{sec:conclusion}
In this work, we present {\tt Deep Horizon}, a combination of two convolutional deep neural networks that can recover input parameters of an image of an accreting SMBH. We create realistic mock observations and use these to show that our network can accurately recover the six parameters investigated in this study if we ignore limited telescope resolutions. We show that the current resolution of the Event Horizon Telescope is insufficient to determine all parameters of this study accurately, but is still sufficient to recover the mass and mass accretion rate accurately and could, therefore, confirm the results found by the Event Horizon Telescope collaboration. With future improvements to the resolution of images of black hole shadows, {\tt Deep Horizon} would be able to recover more parameters. We investigated the case of space-based VLBI, which resulted in highly accurate parameter estimations. 

Overall, the proof of concept presented in this paper shows that machine learning is an interesting parameter extracting tool for horizon scale observations that can be of great value for future tests of GR.

\section{Acknowledgments}

The authors thank S. Caron, B. Stienen, C.F. Gammie, J. Lin, M. Johnson, and L. Rezzolla for valuable discussions and feedback during the project, and the two anonymous referees for their constructive comments on our manuscript. This work was funded by the ERC Synergy Grant
``BlackHoleCam-Imaging the Event Horizon of Black Holes'' (Grant 610058,
\cite{Goddi2017}). The Simons Foundation supports the Flatiron Institute. The GRMHD simulations were performed on the {\tt LOEWE} cluster in CSC in Frankfurt, and the ray-tracing simulations on {\tt COMA} in Nijmegen. This research has made use of NASA's Astrophysics Data System. The results and analyses presented in this manuscript was done with the use of the following software: {\tt python} \citep{travis2007,jarrod2011}, {\tt scipy} \citep{jones2001}, {\tt numpy} \citep{walt2011}, and {\tt matplotlib} \citep{hunter2007}.

\end{document}